\documentclass[aps,twocolumn,showpacs,english]{revtex4}

\usepackage[T1]{fontenc}
\usepackage[latin9]{inputenc}
\usepackage{textcomp}
\usepackage{amsmath}
\usepackage{graphicx}
\usepackage{amssymb}
\usepackage{babel}

\begin{document}

\title{Subnanosecond spectral diffusion of a single quantum dot in a nanowire}

\author{G. Sallen$^{1}$, A. Tribu$^{2}$, T. Aichele$^{3}$, R. Andr\'{e}$^{1}$, L. Besombes$^{1}$, \\
C. Bougerol$^{1}$, M. Richard$^{1}$, S. Tatarenko$^{1}$, K. Kheng$^{2}$, and J.-Ph.~Poizat$^{1}$}

\affiliation{$^1$ CEA-CNRS-UJF group 'Nanophysique et Semiconducteurs',
 Institut N\'{e}el, CNRS - Universit\'{e} Joseph Fourier, 38042 Grenoble, France, \\
$^2$ CEA-CNRS-UJF group 'Nanophysique et Semiconducteurs', CEA/INAC/SP2M, 38054 Grenoble, France\\
$^3$ Physics Institute, Humboldt University, Berlin, Germany}

\begin{abstract}
We have studied spectral diffusion of the photoluminescence of a single CdSe quantum dot inserted in a ZnSe nanowire.
We have measured the characteristic diffusion time as a function of pumping power and temperature using  a recently developed technique [G. Sallen et al, Nature Photon. \textbf{4}, 696 (2010)] that offers
 subnanosecond resolution. These data are consistent with a model where only a \emph{single} carrier wanders around in traps located in  the vicinity of the quantum dot.

\end{abstract}

\pacs{42.50.Lc, 78.67.Lt, 78.55.Et}

\maketitle






Elementary light emitters made of a single quantum object are very sensitive to their local environment. Their optical properties such as intensity, polarization or spectrum can be modified by charge or spin changes in their environment.
Spectral diffusion (SD) of a single light emitter corresponds to random spectral jumps of a narrow line within a broader spectral profile.
This effect has been observed in single molecule \cite{Ambrose,Plakhotnik} or single semiconductor quantum dot \cite{Empedocles,Empedocles97,Robinson,Seufert,Turck,Besombes,Besombes2} experiments.
It is generally due to Stark effect caused by the electric field of randomly trapped charges \cite{Empedocles97,Robinson,Seufert,Turck,Besombes}
or  spin fluctuations  \cite{Besombes2} in the vicinity of the emitter.

Single light emitters are very promising objects for quantum information as single photon sources, flying/solid qubit interface \cite{Simon}, or for single photon manipulation \cite{Simon2,Auffeves,GNL}. In quantum cryptography \cite{Beveratos,Waks}, the spectrum of the single photons can be rather broad \cite{Beveratos} and SD  does not significantly affect the performances. On the contrary, quantum logic operation using linear optics as proposed by Knill et al \cite{KLM} or quantum logic gates using nonlinear interaction at the single photon level (see \cite{GNL} and reference therein) require undistinguishable single photons with ideal spectral purity. This means that they must  necessarily be SD-free, which is also the case for quantum memories.
A solution to be SD-free is to operate faster than spectral diffusion.
The knowledge of the characteristic time of spectral diffusion is thus required
to know whether such a solution is possible.

\begin{figure}
\resizebox{0.4\textwidth}{!}{\includegraphics{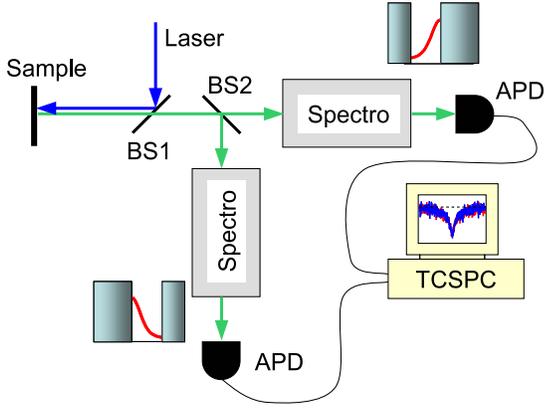}}
 \caption{
 The experimental set-up is based on a standard microphotoluminescence experiment operating at a temperature of T=4K.
 A continuous wave diode laser emitting at $405$ nm excites the sample via beam splitter BS1 with reflection R1=30 \%.
The luminescence  is then transmitted through BS1 and split by  50/50 beamsplitter BS2.
 Each beam is sent to a monochromator (resolution $0.2$ meV) whose output slit illuminates an avalanche photodiode (APD) connected to
a time-correlated single photon (TCSP) module that
builds an histogram of the time delays between  photons.
This allows us to perform either auto-correlation when the two spectrometers (Spectro) are tuned at the same wavelength or cross-correlation otherwise.
The output slits of the spectrometers are adjusted so that a controlled spectral window of a given line is detected.
The work presented below has been obtained with high quantum efficiency APDs  ($\eta=60 \%$ at $550$ nm). With these APDs  the measured timing resolution of the whole set-up is $800$ ps.
}
 \label{fig:exp_setup}
\end{figure}
\begin{figure}
\resizebox{0.4\textwidth}{!}{\includegraphics{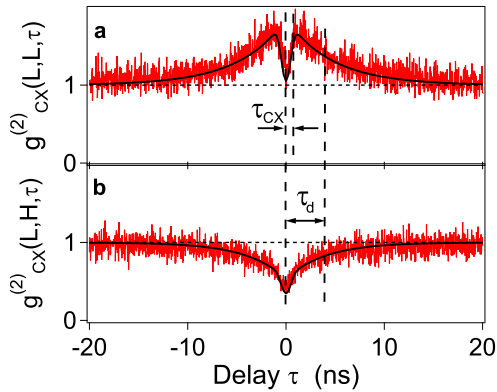}}
 \caption{(a) Auto-correlation of one half of the profile showing the bunching due to SD ($\tau_d=4$ ns) and the narrower single photon antibunching.
 (b), Cross-correlation between the two halves of the profile displaying the antibunching due to SD with the same characteristic time $\tau_d=4$ ns.
 These two plots have been obtained on the same QD with the same excitation power.
 The solid lines are fits with the model explained in \cite{Sallen}.}
 \label{fig:fig1gammafPT}
\end{figure}

 Visualizing directly the spectral wandering
  by recording a time series of spectra
  has been so far the usual method to observe SD \cite{Ambrose,Empedocles,Robinson,Seufert,Turck,Besombes}.
 For single photon emitters, the time resolution was therefore limited by the minimum time of about 1ms required for a
photon counting charged coupled device (CCD) to acquire a spectrum.
Palinginis et al \cite{Palinginis} have improved this resolution by
measuring a modulation frequency-dependent linewidth in a
spectral hole-burning experiment using inhomogeneously
broadened ensembles of semiconducting nanocrystals.
Our recently developed technique \cite{Sallen} converts spectral fluctuations into intensity fluctuations, as also reported in \cite{Plakhotnik,Zumbusch,Coolen,Marshall}. It benefits  from the subnanosecond time resolution of an Hanbury-Brown and Twiss photon correlation set-up and improves by more than 4 orders of magnitude the accessible SD times. It gives access for the first time to SD time in the ns range as shown in this paper. This technique is presented in details in ref \cite{Sallen}. A sketch of the experimental set-up is shown in fig.\ref{fig:exp_setup}. In short, it is based on correlations of photons emitted within a spectral window narrower than the SD broadened line. Owing to the wandering of the homogeneous line, the emission energy stays a limited time within this spectral window leading to photon bunching. The characteristics time of this effect can be easily accessed by photon correlation. fig.\ref{fig:fig1gammafPT}(a) shows a typical result for autocorrelation on one half of the line, and fig.\ref{fig:fig1gammafPT}(b) for cross-correlation between the two  halves of the line.

In the present work we  use this method \cite{Sallen} to evaluate the SD characteristic time of CdSe quantum dots embedded in a ZnSe nanowire \cite{Aichele,Tribu,correlation} and obtain informations about the SD mechanism in this system.

\begin{figure}
\resizebox{0.4\textwidth}{!}{\includegraphics{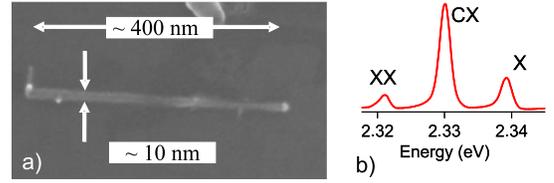}}
 \caption{(a) Scanning electron microscope image of an ZnSe nanowire containing a CdSe quantum dot. (b) Photoluminescence spectrum.}
 \label{fig:sample}
\end{figure}

Semiconductor nanowires (NWs) appear as  promising building blocks for nanoscale devices and circuits since
they can be grown almost  defect free  on low-cost, routinely used substrates such as silicon \cite{Li}.
Furthermore,  NW heterostructures are much less limited by lattice mismatches which greatly widens the possible materials combinations compared to standard self-assembled quantum dots (QDs).
Details on the growth of the CdSe/ZnSe NWs can be found in  \cite{Aichele}.
Their diameter is around $10$nm.
An image of the sample is shown in fig. \ref{fig:sample} together with a typical photoluminescence spectrum.
Exciton (X), biexciton (XX) and charged exciton (CX) lines have been identified unambiguously using photon correlation spectroscopy \cite{correlation}.  According to previously published results \cite{Turck_PSSB,Patton}, it is commonly admitted that charged excitons in CdSe/ZnSe QDs are negatively charged.
The radiative lifetimes of these transitions  are respectively $\tau_X=700$ ps, $\tau_{XX}=400$ ps, $\tau_{CX}=600$ ps. The luminescence wavelength is around $550$ nm with a high count rate of $25 000$ counts per second at $T=4$K. This system has demonstrated single photon generation up to a temperature of $T=220$K \cite{Tribu}.
The QD is either neutral (X and XX lines) or charged (CX line). It can be seen in fig. \ref{fig:sample}(b) that the CX line is more intense than the X and XX lines, indicating that the QD spends more time in the charged state than in the neutral states \cite{correlation}.


The SD rate is  extracted directly from the time width of the correlation measurement as shown in fig. \ref{fig:fig1gammafPT}.
In practice, a more accurate value of the diffusion rate $\gamma_d$ is obtained by
fitting  the antibunching trace coming from cross-correlation data  (fig. \ref{fig:fig1gammafPT}(b)) rather than the fitting from the autocorrelation data (fig. \ref{fig:fig1gammafPT}(a)) since the latter contains the excitonic lifetime as an extra time scale. Details of the model used for the fitting are given in \cite{Sallen}.

Systematic cross-correlation data analysis allowed us to  study the SD rate as a function of different parameters. We have plotted in fig \ref{fig:diffusion(P,T)}(a)
the diffusion rate $\gamma_d$ as a function of pumping power at two different temperatures.
It can be seen that the diffusion rate increases as the pumping power is raised. Higher temperatures leads also to larger diffusion rates.

To explain the power and temperature dependency shown in fig.\ref{fig:diffusion(P,T)}, we have in mind a model with charge traps in the vicinity of the QD \cite{Berthelot}. Electrical charges are randomly trapped and induce a fluctuating electric field leading to spectral diffusion of  QD transitions via  Stark effect.
To account for the observed SD broadened linewidth,
charge traps need to be
located within a few nm of the QD  \cite{Empedocles97,Robinson}.
 In the case of ZnSe the residual doping is of n-type and we assume that the charges are electrons.
Surprisingly, we have observed that neither  the SD-induced linewidth (i.e amplitude of the spectral fluctuations) nor the  lineshape  depend on the pumping power, as shown in fig.\ref{fig:diffusion(P,T)}(b). A possible explanation would be that  there is only room for a \emph{single} charge exploring several trapping sites around the QD and that additional  charges are blocked by Coulomb repulsion. Indeed the presence of several charges, that could be induced by large pump power, would lead to a broadening of the SD-induced linewidth, which is not observed.

The charge number $c$ trapped around the QD is governed  by the following rate equation :
\begin{equation}
\frac{dc}{dt} =\gamma'_{in}(T) (N+N_o)(1-c)-\gamma'_{out}(T) N c ,
\end{equation}
where the first (second) term is the loading (escape) term. The loading term is proportional to $(1-c)$ to account for the fact that the maximum number of charge is one. The loading rate $\gamma'_{in}(T) (N+N_o)$ is proportional to the number  of electrons $N+N_o$ in the ZnSe barrier, where $N$ corresponds to the number of photocreated electron-hole pairs in the barrier and $N_o$ to the residual doping.  For the escape mechanism, we assume that the prominent effect is the
recombination of the trapped electron with a photocreated hole.
This leads to an escape rate $\gamma'_{out}(T) N$ also proportional to $N$.
We assume an activation type behavior  $\sim \exp(-E_a/kT)$  for the temperature dependency of $\gamma'_{in}(T)$ and $\gamma'_{out}(T)$ \cite{Kamada}. The energy $E_a$ corresponds to  shallow potential fluctuations experienced by the charges in the barrier.  Raising the temperature increases their diffusion length and  makes it more likely for the electrons (holes) to load (empty) the deeper traps causing the spectral diffusion of the QD.
From the increase of spectral diffusion with temperature, $\gamma_d (10K)/\gamma_d (4K)= 1.85 \pm 0.15$ at low powers, we can infer an activation energy  $E_a=0.35 \pm 0.05$ meV.

Within this model, the spectral diffusion rate $\gamma_d$ is  given by
\begin{equation}
\gamma_d=\gamma'_{in}(T) N_o + (\gamma'_{in}(T)+\gamma'_{out}(T)) N.
\label{Gd}
\end{equation}
From the experimental  data displayed in fig.\ref{fig:diffusion(P,T)}, it appears that the diffusion rate approaches zero as the pumping power vanishes. This means that the contribution of  $\gamma'_{in}(T) N_o$ is negligible compared to the term depending on the carrier number $N$.
As seen in fig \ref{fig:diffusion(P,T)}(a) the diffusion rates at T=4K and T=10K exhibit a sublinear power dependency. So does the total  light intensity $\Sigma=X+XX+CX$ emitted within  the exciton, biexciton and charged exciton lines. The quantity $\Sigma$ is proportional to the carrier number $N$ in  the nanowire as long as  these three lines are not saturated, which is the case up to a pump power of $4\mu$W in our situation. For larger pump power, higher order multiexcitonic lines appear away from the detected spectral window, and the $\Sigma$ power dependency slightly underestimates the total amount of carrier in the nanowire. The similar power dependencies of $\gamma_d$ and $\Sigma$ is therefore a good indication of the validity of equation (\ref{Gd}) but it is not an unambiguous proof.
 Additionally we mention that sub-linear power behavior of the total amount of carrier could also partly be attributed to Auger scattering in the barrier \cite{OHara, Berthelot}.

We compare now the spectral diffusion rate with the charged to neutral hopping rate $\gamma_C$ also displayed in fig.\ref{fig:diffusion(P,T)}. This rate has been extracted from the width of the bunching peak of the autocorrelation of the whole line of the charged exciton \cite{correlation}.
 The similar hopping rate $\gamma_C$ power dependency  suggests a charged/neutral hopping mechanism similar to the spectral diffusion one as exposed above \cite{Baier}.
 The comparison between the absolute quantitative values of $\gamma_C$ and $\gamma_d$ is not fully reliable since the data have been acquired on different days. Nevertheless the close values of $\gamma_C$ and $\gamma_d$ are in favor of a scenario in which a spectral diffusion event is a jump into the other charge state and back with a modified surrounding charge distribution.

\begin{figure}
\resizebox{0.45\textwidth}{!}{\includegraphics{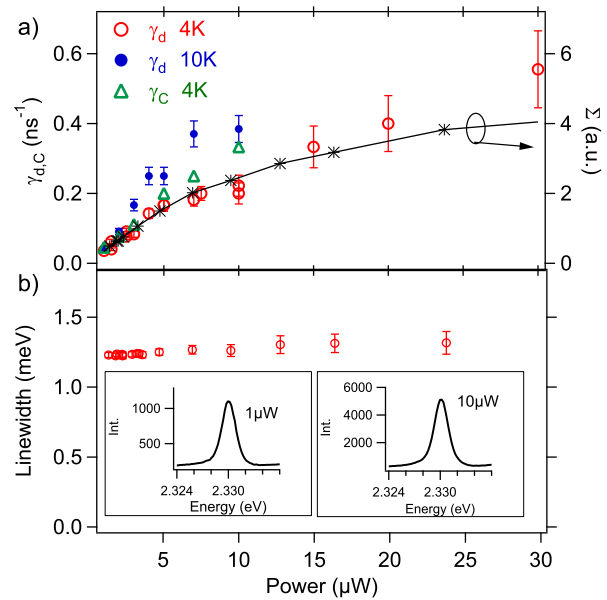}}
 \caption{ a) Left axis : Diffusion rate $\gamma_d$  at T=4K (empty red circles), T=10K (filled blue circles) and the charged to neutral hopping rate $\gamma_C$ (empty green triangles,  see text)  as a function of  exciting power.
 Right axis : The black stars  are proportional to the total amount of light $\Sigma$ emitted by X, CX and XX, and a fair approximation of the total number of carrier $N$ in the NW.
 The saturation pumping power for the charged exciton is $4\mu W$.
 b) Corresponding linewidth as a function of  exciting power at T=4K. Insets : example of two spectra taken at different powers showing that not only the linewidth but also the shape is unchanged.}
 \label{fig:diffusion(P,T)}
\end{figure}

To summarize, we have used a recently developed method \cite{Sallen} to measure spectral diffusion of single emitters with a subnanosecond resolution.  We have used this technique to study, as a function of pumping power and temperature, the spectral diffusion characteristics of the photoluminescence of a single CdSe quantum dot  inserted in a ZnSe nanowire. We have found a characteristic SD time $\tau_d$ of a few  ns and have given good indications that this rate $\gamma_d=1/\tau_d$ is proportional to the carrier number in the wire. In addition to the power independent linewidth, these findings support a model where a \emph{single} charge is wandering within a few nm around the quantum dot, suggesting
a spectral diffusion mechanism closely related to the charged/neutral QD  hopping events.

T.A. acknowledges support by Deutscher
Akademischer Austauschdienst (DAAD). Part of this
work was supported by European project QAP (Contract No.
15848).


\begin{thebibliography}{10}





\bibitem{Ambrose}  W.P. Ambrose, and W.E. Moerner,
Nature \textbf{349}, 225 (1991).

\bibitem{Plakhotnik} T. Plakhotnik and D. Walser, Phys. Rev. Lett. 80, 4064 (1998)

\bibitem{Empedocles} S.A. Empedocles, D.J. Norris, and M.G. Bawendi,
Phys. Rev. Lett. \textbf{77}, 3873 (1996).

\bibitem{Robinson} H.D. Robinson and B.B. Goldberg,
Phys. Rev. B  \textbf{61},  R5086 (2000).

\bibitem{Seufert}
J. Seufert, R. Weigand, G. Bacher, T. Kümmell, A. Forchel, K. Leonardi, and D. Hommel
Appl. Phys. Lett. \textbf{76}, 1872 (2000)

\bibitem{Turck}
V. T\"{u}rck, S. Rodt, O. Stier, R. Heitz, R. Engelhardt, U.W. Pohl, and D. Bimberg,
Phys. Rev. B  \textbf{61}, 9944 (2000).

\bibitem{Besombes}
L. Besombes, K. Kheng, L. Marsal, and H. Mariette,
Phys. Rev. B \textbf{65}, 121314 (2002).

\bibitem{Empedocles97} S.A. Empedocles and M.G. Bawendi, Science  \textbf{278}, 2114 (1997).

\bibitem{Besombes2}
L. Besombes, Y. Leger, J. Bernos, H. Boukari, H. Mariette, J. P. Poizat, T. Clement, J. Fernández-Rossier, and R. Aguado, Phys. Rev. B \textbf{78}, 125324 (2008)

\bibitem{Simon} C. Simon et al, Eur Phys Journal D \textbf{58}, 1 (2010)

\bibitem{Simon2} C. Simon, Y.-M. Niquet, X. Caillet, J. Eymery, J.-P. Poizat, and J.-M. Gérard,
Phys. Rev. B \textbf{75}, 081302 (2007)

\bibitem{Auffeves}
A. Auffèves-Garnier, C. Simon, J.-M. Gérard, and J.P. Poizat,
Phys. Rev. A 75, 053823 (2007)

\bibitem{GNL}
J.L. O'Brien, A. Furusawa, and J. Vu\u{c}kovi\'{c},
Nature Photon. \textbf{3}, 687 (2009)

\bibitem{Beveratos}
A. Beveratos, R. Brouri, T. Gacoin, A. Villing, J.P. Poizat, and P. Grangier
Phys. Rev. Lett. 89, 187901 (2002)

\bibitem{Waks}
E. Waks, K. Inoue, C. Santori, D. Fattal, J. Vuckovic, G.S. Solomon, and Y. Yamamoto
Nature \textbf{420}, 762 (2002)

\bibitem{KLM}
E. Knill, R. Laflamme, G. J. Milburn, Nature \textbf{409}, 46 (2001); see also
A. Kiraz, M. Atat\"{u}re, and A. Imamo\u{g}lu, Phys. Rev. A 69, 032305 (2004) and
P. Kok, W. J. Munro, K. Nemoto, T. C. Ralph, J.P. Dowling, and G. J. Milburn
Rev. Mod. Phys. \textbf{79}, 135 (2007)

\bibitem{Palinginis} P. Palinginis, S. Tavenner, M. Lonergan, and H. Wang
Phys. Rev. B \textbf{67}, 201307 (2003).

\bibitem{Sallen} G. Sallen, A. Tribu, T. Aichele, R. Andr\'{e}, L. Besombes, C. Bougerol, M. Richard, S. Tatarenko, K. Kheng, and J. Ph.~Poizat, Nature Photon. \textbf{4}, 696 (2010)
 	



\bibitem{Zumbusch} Zumbusch, A., Fleury, L.,  Brown, R.,  Bernard, J. \&  Orrit, M.
Probing individual two-level systems in a polymer by correlations of single molecule fluorescence,
Phys. Rev. Lett. \textbf{70}, 3584-3587 (1993).


\bibitem{Coolen}
L. Coolen, X. Brokmann, P. Spinicelli, and J.-P. Hermier
Phys. Rev. Lett. \textbf{100}, 027403 (2008).


\bibitem{Marshall}
L.F. Marshall, Jian Cui, X. Brokmann, and M.G. Bawendi,
Phys. Rev. Lett. \textbf{105}, 053005 (2010)


\bibitem{Aichele}
T. Aichele, A. Tribu, C. Bougerol, K. Kheng, R. Andr\'{e}, and S. Tatarenko
Appl. Phys. Lett. \textbf{93}, 143106 (2008).

\bibitem{correlation} G. Sallen, A. Tribu, T. Aichele, R. Andr\'{e}, L. Besombes, C. Bougerol, S. Tatarenko, K. Kheng, and J. Ph.~Poizat, Phys. Rev. B  \textbf{80}, 085310 (2009).

\bibitem{Tribu} A. Tribu,  G. Sallen, T. Aichele, R Andr\'{e}, J.-Ph.
Poizat, C. Bougerol, S. Tatarenko, and K. Kheng, Nano Lett.  \textbf{8}, 4326 (2008).

\bibitem{Turck_PSSB} V. Türck, S. Rodt, R. Heitz, O. Stier, M. Strassburg, U. W. Pohl,
and D. Bimberg, Phys. Status Solidi B \textbf{224}, 217 (2001).

\bibitem{Patton} B. Patton, W. Langbein, and U. Woggon, Phys. Rev. B \textbf{68},
125316 (2003).


\bibitem{Li} Y. Li, F. Qian, J. Xiang, and C.M. Lieber, Materials today 9, 18 (2006)

\bibitem{Berthelot} A. Berthelot, I. Favero, G. Cassabois, C. Voisin, C. Delalande, Ph. Roussignol, R. Ferreira,
and J.M. G\'{e}rard,
Nature Physics \textbf{2},  759  (2006).

\bibitem{OHara}K. E. O'Hara, J. R. Gullingsrud, and J. P. Wolfe , Phys. Rev. B \textbf{60}, 10872 (1999)



\bibitem{Kamada}  H. Kamada and T. Kutsuwa
Phys. Rev. B \textbf{78}, 155324 (2008)

\bibitem{Baier} M. H. Baier, A. Malko, E. Pelucchi, D. Y. Oberli, and E. Kapon
Phys. Rev. B \textbf{73}, 205321 (2006)




\bibitem{Finley}
J.J. Finley, M. Sabathil, P. Vogl,  G. Abstreiter,
R. Oulton, A. I. Tartakovskii, D. J. Mowbray,  M. S. Skolnick,
S. L. Liew, A. G. Cullis, and M. Hopkinson,
Phys. Rev. B \textbf{70}, 201308(R) (2004)


\end{thebibliography}
\end{document}